\documentstyle[aps,preprint,tighten]{revtex}
\def\G{\Gamma}

\def\w{\Omega}
\def\C{{\bf C}}

\def\la{\langle}
\def\ra{\rangle}
\def\be{\nopagebreak[3]\begin{equation}}
        \def\ee{\end{equation}}
        \def\ba{\nopagebreak[3]\begin{eqnarray}}
        \def\ea{\end{eqnarray}}

\newcommand{\teta}{\rlap{\lower2ex\hbox{$\,\tilde{}$}}\eta{}}

\preprint{\vbox{\baselineskip=12pt
 \rightline{ICN-UNAM-02-02}
 \rightline{gr-qc/0204053}}}
       \begin{document}
       \draft

\title{On Unitary Time Evolution in  Gowdy $T^3$ Cosmologies}
\author{Alejandro Corichi${}^{1,2}$\thanks{Email: corichi, cortez,
quevedo@nuclecu.unam.mx},
Jer\'{o}nimo Cortez${}^1$ and Hernando Quevedo${}^1$ }
\address{1. Instituto de Ciencias Nucleares\\
Universidad Nacional Aut\'onoma de M\'exico\\
A. Postal 70-543, M\'exico D.F. 04510, MEXICO \\
}
\address{2. Department of Physics and Astronomy\\
University of Mississippi,
 University, MS 38677, USA\\
}
\maketitle

\begin{abstract}

A non-perturbative canonical quantization of Gowdy $T^3$ polarized models
carried out recently is considered. This approach
profits from the equivalence between the symmetry reduced model
and $2+1$ gravity coupled to a massless real scalar field. The
system is partially gauge fixed and a choice of internal time is
performed, for which the true degrees of freedom of the model
reduce to a massless free scalar field propagating on a
2-dimensional expanding torus. It is shown that the symplectic
transformation that determines the classical dynamics cannot be
unitarily implemented on the corresponding Hilbert space of
quantum states. The implications of this result for both
quantization of fields on curved manifolds and physically relevant
questions regarding the initial singularity are discussed.

\end{abstract}



\section{Introduction}

In the search for a quantum theory of gravity within the canonical
approach, it has been historically useful to consider symmetry
reduced models. The most studied examples are homogeneous models,
where the infinite dimensional system is reduced to a model with a
finite number of degrees of freedom. These are known as
mini-superspace models \cite{misner1}. Another class of symmetry
reduced models where the resulting system is still a field theory
with an infinite number of degrees of freedom are known as
midi-superspace models (for a recent review see \cite{torre}).
Recently, within this class, the models that have received special
attention are the Einstein-Rosen waves and Gowdy cosmological
models \cite{ashpie,pierri1,pierri2}. One interesting feature of
this type of models is that due to their spacetime symmetries, the
classical dynamics turns out to be derivable from an equivalent
complete integrable system, but they still possess an infinite
number of degrees of freedom so that their quantization would lead
to a true quantum field theory (in contrast to mini-superspace
models with a finite number of degrees of freedom). In this work,
we will consider the canonical quantization of the polarized Gowdy
$T^3$ cosmological model \cite{gowdy}. This is the simplest
inhomogeneous, empty, spatially closed cosmological model. It was
extensively studied in the 70's \cite{berger,misner2}, and
subsequently re-examined by several authors \cite{husain,mena}. Of
particular relevance is the recent work by Pierri, where definite
progress was achieved in defining a rigorous quantization of the
model \cite{pierri1}. In this case, the quantization is based on
the fact that the corresponding gravitational field can be
equivalently treated as $2+1$ gravity coupled to  a massless
scalar field, defined on a $2+1$-dimensional manifold with
topology $T^2\times R$.

One important aspect in the study of quantum cosmological models
is dynamical evolution; that is the dynamics in which a physical
quantum state evolves from an initial Cauchy surface to a final
one.  Recall that in general relativity there are no preferred
foliations in spacetime and the dynamical evolution should
consider all possible spacelike foliations in order to be in
agreement with the requirement of general covariance. Furthermore,
in the case in which the Cauchy surfaces are compact, dynamical
evolution is pure gauge, so any interpretation of time evolution
is normally through a deparametrization procedure which, in the
Hamiltonian language, is normally achieved via ``time dependent
gauge fixing''. Thus, the dynamics to be considered in quantum
cosmological models concerns the evolution of  quantum states
between Cauchy surfaces, defined by the particular gauge choice.
Different choices of time parameters may lead to inequivalent
quantizations. This fact is true even in the simple mechanical
mini-superspace models. In the particular model we are interested,
the system is partially gauge fixed at the classical level, and in
particular a time function $T$ is chosen and interpreted as the
time which defines ``evolution". The surfaces of constant $T$ are
Cauchy surfaces of a fiducial flat background, so the model get
reduced to a quantum scalar field on a flat background, equipped
with a foliation of preferred surfaces which define ``time
evolution" in the corresponding quantum gravitational system.

At the classical level, this dynamical evolution of the field can
be represented as a canonical transformation that acts on points
of the corresponding phase space. The question is whether, at the
quantum level, this canonical transformation can be implemented on
the space of quantum states of the field by means of a unitary
operator. This is a rather delicate problem that has been analyzed
in detail only recently and for a few special cases
\cite{kuchar,helfer,torrvar1,torrvar2}, all of them concerning
free scalar fields on flat or stationary spacetimes. Fortunately,
the quantum polarized Gowdy $T^3$ cosmological models belong to
this class, and so we will be able to investigate the question
about the unitary implementability of these models within this
approach. This is the main goal of the present work.

We will show that for the particular quantization performed in
\cite{pierri1,pierri2}, time evolution is not implementable as a
unitary evolution.
 Given that, at the classical level time evolution is pure gauge,
and the particular choice of time is an ad-hock procedure to regain
dynamics from a purely frozen formalism, one might argue that
unitary implementability of this fictitious time evolution is not
necessary for the consistency of the formalism. However, as we
will argue, the implementability is needed in order to ask
physically meaningful questions regarding, say, the initial
singularity. That is, questions such as whether the initial
singularity is smeared by quantum effects should have a definite
answer within a consistent quantization. Thus, we shall conclude
that we can not extract any physics out of these models as
presently constructed.

This paper is organized as follows. In Sec.~\ref{sec:2} we review
the quantization of polarized Gowdy $T^3$ cosmological models as
performed by Pierri \cite{pierri1,pierri2}. We show that the two
different sets of creation
 and annihilation operators proposed in this quantization are related
by means of a unitary Bogoliubov  transformation. In
Sec.~\ref{sec:3.a} we explicitly calculate the canonical
(symplectic) transformation that represents the classical
evolution of the system. In Sec.~\ref{sec:3.b} we prove that this
canonical transformation is not unitarily implementable on the
corresponding Fock space. We end with a discussion and some
conclusions in Sec.~\ref{sec:5}.

\section{Canonical quantization}
\label{sec:2}

 The polarized Gowdy $T^3$ models are globally hyperbolic
four-dimensional vacuum spacetimes, with two commuting hypersurface
orthogonal spacelike Killing fields and compact spacelike
hypersurfaces homeomorphic to a three-torus. Because this system can 
be equivalent treated as 2+1 gravity (minimally) coupled to an 
axi-symmetric massless scalar field, let us begin by considering the action
\be
 S\bigl[ \, ^{{\mbox{\tiny{(3)}}}}g,\psi \, \bigr]={{1}\over{2 \pi}}
\int_{^{{\mbox{\tiny{(3)}}}}M}d^{3}x \sqrt{-{^{{\mbox{\tiny{(3)}}}}g}}
 \ \bigl(
{^{{\mbox{\tiny{(3)}}}}R}-{^{{\mbox{\tiny{(3)}}}}g^{ab}} \nabla_{a}\psi 
\nabla_{b}\psi\bigr)
\label{action}
\ee
where $^{{\mbox{\tiny{(3)}}}}R$ is the Ricci scalar of the
$3$-d spacetime $(^{{\mbox{\tiny{(3)}}}}M,^{{\mbox{\tiny{(3)}}}}g_{ab})$,
 $^{{\mbox{\tiny{(3)}}}}M$ is a $3$-d manifold with topology $T^{2} \times
 \bf{R}$ and spacetime metric  $^{{\mbox{\tiny{(3)}}}}g_{ab}=h_{ab}+
\tau^{2}\nabla_{a}\sigma\nabla_{b}\sigma$. The Killing field
$\sigma^{a}$ is hypersurface orthogonal and the field $h_{ab}$ is
a metric of signature $(-,+)$ on the $2$-manifold orthogonal to
$\sigma^{a}$; $\tau$ is the norm of $\sigma^{a}$ and $\sigma$ is
an angular coordinate with range $0\leq \sigma < 2\pi$ such
 that $\sigma^{a}\nabla_{a}\sigma =1$.

Introducing a generic slicing
by compact spacelike hypersurfaces labeled by $t=const$ the
$2$-metric can be written as $h_{ab}=\bigl(
-N^{2}+N^{\theta}N_{\theta}\bigr)\nabla_{a}t\nabla_{b}t+
2N_{\theta}\nabla_{(a}t\nabla_{b)}\theta+e^{\gamma}
\nabla_{a}\theta\nabla_{b}\theta$, where the lapse, $N$, the
shift, $N^{\theta}$, and $\gamma$ are functions of $\theta$ and
$t$. The angular coordinate $\theta \in [0,2\pi)$ is such that
$\hat{\theta}^{a}\nabla_{a}\theta=1$, where $\hat{\theta}^{a}$ is
the unit vector field within each slice orthogonal to
$\sigma^{a}$.
 Thus the system consists of five functions $(N,\ N^{\theta},
\ \gamma,\ \tau,\ \psi)$ of $t$ and $\theta$ which are periodic in
$\theta$. The function $\psi$ represents the zero rest mass scalar
field.

Substituting the expression for $^{{\mbox{\tiny{(3)}}}}g_{ab}$ in
the action
(\ref{action}) we pass to the Hamiltonian formulation
\begin{equation}
\label{hamil-action} S=\int dt \biggl( \oint  \bigl(
p_{\gamma}\dot{\gamma}+ p_{\tau}\dot{\tau}+ p_{\psi}\dot{\psi}
\bigr)\biggr)-H[N,N^{\theta}]
\end{equation}
where the Hamiltonian $H$ is given by $H[N,N^{\theta}]=\oint
\bigl( NC+N^{\theta}C_{\theta}\bigr)$ (here, the symbol $\oint$ 
denotes integration over $\theta \in S^{1}$) and the first class
constraints $C$ and $C^{\theta}$ are
\begin{eqnarray}
\label{hamil-constraints}
C=e^{-\gamma / 2}\biggl[ 2 \tau '' -\gamma'\tau'-p_{\gamma} p_{\tau} +
\tau \biggl({{p_{\psi}^{2}}\over{4\tau ^{2}}}+\psi'^{2} \biggr) \biggr] \\
C^{\theta}=e^{-\gamma}\bigl(\gamma'p_{\gamma}-2p_{\gamma}'+\tau'p_{\tau}
 +p_{\psi}\psi' \bigr)
\end{eqnarray}
The lapse and shift are not dynamical variables, thus the
phase space $\G$ consists of three canonically-conjugate pairs
of periodic functions of $\theta$, $(\gamma,p_{\gamma};
\tau,p_{\tau};\psi , p_{\psi})$ on a 2-d manifold $\Sigma$
 with topology $T^{2}$.

Because the Hamiltonian vanishes on the constraint surface there is
no distinction between gauge and dynamics and therefore it is necessary to
introduce a ``deparametrization'' procedure to discuss dynamics.
From the infinite set of vector fields generated by the
Hamiltonian constraints we select one to represent evolution and
gauge fix the others. For gauge fixing let us demand
\begin{equation}
\label{coord-condition} p_{\gamma}+p=0 \qquad \mbox{and} \qquad
\tau (\theta)'=0
\end{equation}
where $p$ is a spatial constant that has zero Poisson bracket with
all the constraints and hence it can not be removed by gauge
fixing. The second condition will allow us to regard $\tau
(\theta)$ as the time parameter.

The consistency of the formalism requires that the Poisson
brackets $\{\tau (\theta)',H[N,N^{\theta}]\}$ and
$\{p_{\gamma}+p,H[N,N^{\theta}] \}$ vanish. This can be achieved
if the freely specifiable $N$ and $N^{\theta}$ are chosen as,
\begin{equation}
\label{lapse-shift} N= {{e^{\gamma /2}}\over{p}} \:\:\:\:
\mbox{and} \:\:\:\:
 N^{\theta}=0
\end{equation}
therefore the coordinate condition (\ref{coord-condition}) is
acceptable. Indeed, for the special choice (\ref{lapse-shift}) we have
that $\dot{\tau}(\theta)=1$ and $\dot{p}_\gamma(\theta) = -\dot{p}= 0$ and
hence $p$ becomes a true constant that can be associated with a
time scaling parameter in the original 3+1-spacetime.  
Thus, apart from a global degree of freedom, the true
degrees of freedom will all reside in the field $\psi$. Solving
the set of second class constraints
$(C,C^{\theta},p_{\gamma}+p,\tau ')$ the result is
\begin{equation}
\label{tau-constraint} p_{\tau}=-{{t}\over{p}}\biggl(
{{p_{\psi}^{2}}\over{4t ^{2}}}+ \psi'^{2} \biggr)
\end{equation}
\begin{equation}
\label{gamma-constraint}
\gamma(\theta)={{1}\over{p}}\int_{0}^{\theta}d \theta_{1}p_{\psi}
\psi'+ \gamma(0)
\end{equation}
Since $\gamma$ is a smooth function of $\theta$, it must admit a
Fourier series of the form $\gamma=q+\sum_{n \neq 0} {{e^{in
\theta}}\over{\sqrt{2\pi}}}\gamma_{n}$, then $\sum_{n \neq
0}\bigl({{e^{in \theta}-1}\over{\sqrt{2\pi}}}
\bigr)\gamma_{n}={{1}\over{p}} \int_{0}^{\theta}d
\theta_{1}p_{\psi}\psi'$ and we can solve for all modes
but the zero mode. i.e., we can solve (\ref{gamma-constraint})
for $\bar{\gamma}:=\sum_{n \neq 0} {{e^{in \theta}}
\over{\sqrt{2\pi}}}\gamma_{n}$ and we are left with the global
degree of freedom $q={{1}\over{2\pi}}\oint  \gamma$ unsolved. This
is consistent with the fact that a constant shift vector
$N^{\theta}={\rm const}$ would also be acceptable for preserving the
conditions (\ref{coord-condition}). Substituting
(\ref{gamma-constraint}) in (\ref{lapse-shift}) we obtain that
$N=N[q,p,\psi,p_{\psi}]$. The spacetime metric is now completely
determined by $q$, $p$, $\psi$ and $p_{\psi}$
\begin{equation}
\label{3-metric}
^{{\mbox{\tiny{(3)}}}}g_{ab}=e^{q+\bar{\gamma}}\biggl(-{{1}\over{p^{2}}}
\nabla_{a}t\nabla_{b}t+\nabla_{a}\theta \nabla_{b}\theta \biggr) +
t^{2}\nabla_{a}\sigma \nabla_{b}\sigma
\end{equation}
The phase space variables are periodic functions of $\theta$,
therefore $\gamma(2\pi)-\gamma(0)=0$ defines via
Eq.(\ref{gamma-constraint}) the global constraint
\be
P_{\theta}:=\oint p_\psi\,\psi^\prime=0
\ee

The non-degenerate symplectic structure on the reduced phase space
$\G_{r}=\G_{g} \oplus \bar{\G}$, where
$\G_{g}$ is coordinatized by the pair $(q,p)$
and $\bar{\Gamma}$ by $(\psi,p_{\psi})$, is the pull-back of the
natural symplectic structure defined on $\G$. Thus $\{q,p\}=1$ and
$\{\psi(\theta_{1}),p_{\psi}(\theta_{2})\}=\delta(\theta_{1},
\theta_{2})$ on $\Gamma_{r}$.

Substituting (\ref{coord-condition}) and (\ref{tau-constraint}) in
(\ref{hamil-action}) we obtain the reduced action
\begin{equation}
\label{reduced action} S=\int dt \Biggl ( p \dot{q} + \oint
\biggl[ p_{\psi} \dot{\psi} -{{t}\over{p}}\Biggl(
{{p_{\psi}^{2}}\over{4t^{2}}}+\psi'^{2}\Biggr)\biggr] \Biggr)
\end{equation}
and the reduced Hamiltonian
\begin{equation}
\label{rh} H= \oint {{t}\over{p}}\Biggl[
{{p_{\psi}^{2}}\over{4t^{2}}}+\psi'^{2}\Biggr]
\end{equation}
Varying the action (\ref{reduced action}) with respect to $\psi$
and $p_{\psi}$, the field equations are ${{\partial
\psi}\over{\partial T}} ={{p_{\psi}}\over{2Tp}}$ and ${{\partial
p_{\psi}}\over{\partial T}}= 2Tp \psi''$,  which is equivalent to
the Klein-Gordon equation for the scalar field $\psi$ propagating
on a fictitious flat background
$^{{\mbox{\tiny{(f)}}}}{g}_{ab}=-\nabla_{a}T\nabla_{b}T+\nabla_{a}
\theta \nabla_{b}\theta + T^{2}\nabla_{a}\sigma \nabla_{b}\sigma$,
with the further restriction that the field $\psi$ does not depend
on $\sigma$. Here the constant rescaling $T:=t/p$ has been
considered, in order to simplify the resulting dynamical equation
for $\psi$. Hence the phase space $\bar{\Gamma}$, coordinatized by
$\varphi:=\psi(\theta)$ and $\pi:=p_{\psi}(\theta)$, corresponds
to the symplectic vector space $(V,\Omega_{V})$ of smooth real
solutions to the Klein-Gordon equation
$^{{\mbox{\tiny{(f)}}}}g^{ab}\nabla_a\nabla_b\psi=0$, where the
symplectic structure $\Omega_{V}$ is given by
\begin{equation}
\label{symplectic-structure} \Omega_{V}(\psi_{1},\psi_{2})=\oint T
(\psi_{2}
\partial_{T}\psi_{1}-\psi_{1}\partial_{T}\psi_{2})
\end{equation}

Note that the choice of internal time $T:\G \mapsto {\bf{R}}$, depends
on the point of phase space, in particular, on the global degree
of freedom $p$, making it a $q-$number.
Recall that in this case, one has
deparametrized the system, that is, one has defined a fictitious
time evolution with respect to the number $t$.
Strictly speaking, from the canonical viewpoint
one should choose a particular value $t_0$ of the time parameter
in order to fix once and for all a single Cauchy surface
$\Sigma_0$. This would be the ``frozen formalism" description,
where only the true degrees of freedom are left and the notion of
dynamics has been lost. However, for the purposes of quantization,
it is convenient to exploit this deparametrization since this
allows to complete the quantization in a rigorous fashion. However,
in the quantum theory the role of $T$ is very different, namely
the function $T$ does not have an
a-priory meaning as a spacetime time parameter. At best, one
might expect that if one chooses suitable semi-classical states,
a classical notion of time might arise, which could be then
compared to the function $T$.
We shall come back to the issue of ``frozen formalism vs. fake
dynamics" in the discussion section.

Thus,  the problem of quantization of the true degrees of freedom
in this case reduces to a quantum theory of the massless scalar
field $\psi$ on a fictitious background \cite{pierri1}. There is a
convenient way of writing the solutions of the Klein-Gordon
equation,
\be
\label{kg-sol}
\psi(\theta,T)=\sum_{m \in
{\bf{Z}}}f_{m}(\theta,T)\: A_{m}+ \overline{f_{m}(\theta,T)}\:
\overline{A_{m}}\, , \ee
where the bar denotes complex conjugation, the $A_{m}$'s are
arbitrary constants and $f_{0}(\theta,T)=(1/2)\bigl( \ln T -i\bigr), \
f_{m}(\theta,T)=(1/2)H^{(1)}_{0}(\vert m \vert T)e^{im \theta}$
for $m \neq 0$, with $H^{(1)}_{0}$ the $0$th-order Hankel function
of the first kind.

For the quantization of this system one can use the Fock
procedure, starting from the one-particle Hilbert space ${\cal
H}_0$ for which an appropriate complex structure $J_V$ is needed
that must be compatible with the symplectic structure $\Omega_V$.
It can be shown \cite{pierri1} that in this case the complex
structure can be chosen as,
\be
J_{V}(a \ln T):=-a\qquad{\rm and}\qquad J_{V}(a):=a
\ln T,
\ee
for $m= 0$ and as,
\be
J_{V}\bigl( J_{0}(\vert m \vert T) \bigr):=N_{0}(\vert m
\vert T)\quad {\rm and}\quad J_{V}\bigl( N_{0}(\vert m \vert T)
\bigr):=-J_{0}(\vert m \vert T)\,, \ee
for $m \neq 0$, where $a$
is a constant. The (fiducial) Hilbert space can be represented as ${\cal F} =
{\cal H}_g \otimes \bar{ {\cal F}}$, where ${\cal H}_g$ is the
Hilbert space in which the operators $\hat{q}$ and $\hat{p}$ are
well defined, and $ \bar{ {\cal F}}$ is the symmetric Fock space
(associated to the ``one-particle'' Hilbert space ${\cal H}_0$) in
which the field operator
$\hat{\psi}$ can be written as, \be
\hat{\psi}(\theta ,
T)=\sum_{m \in {\bf{Z}}}f_{m}(\theta,T)\: \hat{A}_{m}+
\overline{f_{m}(\theta,T)}\: \hat{A}^{\dag}_{m}\, ,
\ee
in terms of the creation and annihilation operators. The space of
physical states, ${\cal F}_p$, is the subspace of ${\cal F}$ defined by
\be
:\hat{P}_{\theta}:\vert \Psi\rangle_{p}=2 \sum_{m \in \;{\bf{Z}}} m
\hat{A}^{\dag}_{m}\hat{A}_{m} \vert \Psi\rangle_{p}=0\, .
\ee
Finally, the
Hamilton operator can be expressed as,
\be
\hat H  =  {\pi \over
2T}:\hat p (\hat A _0 + \hat A _0^\dagger)^2 : + \sum_{n \in\;
{\bf{Z}}} \hat p ( \alpha_n \hat A _n \hat A _{-n} +
\overline{\alpha_n} \hat A _n^\dagger \hat A _{-n}^\dagger +
2\beta_n \hat A _n^\dagger \hat A _n ) \ee
where $\alpha_n(T) =
(T/4) n^2 [(H_0^{(1)})^2 + (H_1^{(1)})^2]$ and $\beta_n(T) = (T/4)
n^2 [H_1^{(1)}H_1^{(2)} + H_0^{(1)} H_0^{(2)}]$ for $n\neq 0$.
Although this Hamiltonian leaves ${\cal F}_p$ invariant, it has
the disadvantage that the vacuum state is not an eigenvector of it
with zero eigenvalue. In order to avoid this difficulty, a new set
of creation and annihilation operators has been recently proposed
\cite{pierri2}:
\be
\hat a _0  =   {i \over \sqrt{2}}(\sqrt{3}\hat
A _0 + \hat A _0^\dagger),
\ee
and
\be
\hat a _n  =  \sqrt{\pi\over 2}
( \tilde\alpha _n \hat A _n + \tilde\beta _n \hat A_{-n}^\dagger)\, ,
\ee
where $\tilde\alpha _n = \alpha_n /(|n|\tilde\beta _n)$ and
$\tilde\beta _n = \sqrt{(\beta_n -|n|/\pi)/|n|}$. It should be
noted that our choice differs from that in \cite{pierri2}, which
does not satisfy the relations
$[\hat{a}_n,\hat{a}_{n^\prime}^\dagger]=\delta_{n,n^\prime}$. In
this new set the Hamilton operator can be written as,
\be
:\hat H : = {{1}\over{T}}\hat p \,\hat P _0^2
+{{2}\over{\pi}} \sum_{n\in {\bf Z}}|n|\, \hat p\,
\hat a _n^\dagger \hat a_n\, ,
\ee
with $\hat P _0 = i\sqrt{\pi}/(1+\sqrt{3}) (\hat a _0^\dagger
- \hat a _0)$. For this Hamiltonian, the vacuum state is in fact
an eigenvector with zero eigenvalue.
The question arises whether
locally the two different sets of creation and annihilation
operators lead to different quantizations. If so, one would need
to investigate the problem of unitary implementability (to be
treated in the next section) for both sets separately. To answer
this question we have analyzed the coefficients that relate the
old operators $\hat A _n$ with the new ones $\hat a _n$. A
straightforward calculation shows that they satisfy the
relationships,
\be
\sum_k (\tilde \alpha _{mk} \overline{ \tilde
\alpha}_{nk} - \tilde \beta _{mk} \overline{ \tilde \beta}_{nk}) =
\delta_{m,n}\, , \ee
and
\be
\sum_k (\tilde \alpha _{mk} { \tilde \beta}_{nk} - \tilde
\beta _{mk} { \tilde \alpha}_{nk}) = 0\, ,
\ee
where $\tilde \alpha _{mk}=\tilde \alpha _{m} \delta_{m,k}$ and 
$\tilde \beta _{mk}=\tilde \beta _{m} \delta_{-m,k}$. Moreover, 
the coefficients $\tilde \beta _{mk}$ are square-summable. 
To see this, let $C(\vert m \vert T) = (\beta_{m}- 
\vert m \vert /\pi)/\vert m \vert$ and let $N$ be a 
positive integer such that $NT >> 1$, then
\be
\sum_{m,n}\vert \tilde \beta_{nm}\vert ^{2} = \sum_{m} \vert \tilde 
\beta_{m} \vert ^{2} = 2 \sum_{m=1}^{N-1}C(mT)+2 \sum_{m=N}^{\infty}C(mT)
\ee
and the condition that the coefficients $\tilde \beta _{mk}$ are 
square-summable is equivalent to 
\be
\label{sq-sum}
\sum_{m=N}^{\infty}C(mT)< \infty
\ee
Now, from the expansions for $J_{n}(x)$ and $N_{n}(x)$ \cite{arfken}; i.e.,
\ba
J_{n}(x)=\sqrt{{{2}\over{\pi x}}} 
\Biggl[P_{n}(x)\cos\left(x-{{(n+{{1}\over{2}})\pi} \over {2}}
\right)-Q_{n}(x)\sin\left(x-{{(n+{{1}\over{2}})\pi} \over {2}}\right)\Biggr] 
\: , 
\\ N_{n}(x)= \sqrt{{{2}\over{\pi x}}} 
\Biggl[P_{n}(x)\sin\left(x-{{(n+{{1}\over{2}})\pi} 
\over {2}}\right)+Q_{n}(x)\cos\left( x-{{(n+{{1}\over{2}})\pi}
 \over {2}}\right)\Biggr] \: , 
\nonumber
\ea
where
\ba
P_{n}(x)&=&1-{{(4n^{2}-1)(4n^{2}-9)}\over
{2!(8x)^{2}}}+{{(4n^{2}-1)(4n^{2}-9)(4n^{2}-25)
(4n^{2}-49)}\over{4!(8x)^{4}}}-\ldots \: , \\ 
Q_{n}(x)&=& {{(4n^{2}-1)}\over{1!(8x)}}-{{(4n^{2}-1)
(4n^{2}-9)(4n^{2}-25)}
\over{3!(8x)^{3}}}+\ldots \: , 
\nonumber
\ea
it is easy to see that 
\be
C(kT)={{Tk}\over{4}}\Biggl[ {{2}\over{\pi Tk}}\, 
(P_{1}^{2}(kT)+P_{0}^{2}(kT)+Q_{1}^{2}(kT)+Q_{0}^{2}(kT))\Biggr]-
{{1}\over{\pi}} .
\ee
That is, for $kT>> 1$ we obtain that
\be
\label{sq-sum-ord}
C(kT)={{Tk}\over{4}}\Biggl[ {{2}\over{\pi Tk}}\, 
(2+O(1/(kT)^{2}))\Biggr]-{{1}\over{\pi}}
\ee
where $O(1/(kT)^{2})$ contains all the terms of the form 
${{c_{i}}\over{(kT)^{n}}}$, 
with $c_{i}$ some (real) constants and ${\bf{N}} \owns n \geq 2$. 
Because $\sum_{k}{{c_{i}}\over{(kT)^{n}}}$ converges (the Riemann 
zeta function defined 
by $\zeta(p)=\sum_{k=1}^{\infty}k^{-p}$ is divergent for $p \leq 1$ 
and convergent for $p>1$) 
then Eq.~(\ref{sq-sum-ord}) implies Eq.~(\ref{sq-sum}) and therefore 
the coefficients 
$\tilde \beta_{mk}$ are square-summable. Thus, the transformation 
between the two different 
sets of creation and annihilation operators is a unitary Bogoliubov 
transformation 
\cite{birdav,honegger}. Consequently, the second set of operators, 
$\hat a _n$, corresponds 
only to the choice of a second complete set of modes $\tilde f _n$. 
Thus, we can
equivalently choose any of these two sets of creation and
annihilation operators given above. We will perform the analysis
of the quantum implementability of the classical dynamical
evolution in Sec.~\ref{sec:3},  using the operators $\hat A _n$ and the
complex structure given above.

\section{Functional Evolution}
\label{sec:3}

It is known that for a massless, free, real scalar field propagating on
a $(n+1)$-dimensional static spacetime with topology
$T^{n}\times{\bf{R}}$, dynamical evolution along arbitrary
spacelike foliations is unitarily implemented on the same Fock
space as that associated with inertial foliations if $n=1$
\cite{torrvar1} and will not be, in general, unitarily
implemented if $n>1$ \cite{torrvar2}. However, for the
(special) case in which we consider time evolution of the
Klein-Gordon field between any two flat spacelike Cauchy
surfaces, dynamical evolution is unitarily implementable for
all positive integers $n$. In this section we will see that
this result actually does not extend to our case, where the
spatial slices have the same topology (tori) but now they are expanding.

\subsection{The symplectic transformation of classical dynamics}
\label{sec:3.a}

It is generally known that the phase space of a real, linear
Klein-Gordon field propagating on a globally hyperbolic background
spacetime $(M\simeq \Sigma \times {\bf R},g_{ab})$ with $\Sigma$ a
compact spacelike Cauchy surface, can be alternatively described by
 the space $\G$ of Cauchy data, that is $\{(\varphi , \pi)\vert \,
\varphi : \Sigma \to {\bf{R}},\, \pi: \Sigma \to {\bf{R}};\varphi ,
 \pi \in C^{\infty}(\Sigma)\}$, or by the space $V$ of smooth
solutions to the Klein-Gordon equation which arises from initial
 data on $\G$ \cite{Ash-Bomb-Reul-wald-books}. Given an embedding
 $E$ of $\Sigma$ as a Cauchy surface $E(\Sigma)$ in $M$, there is
 a natural isomorphism $I_{E}:\G\to V$, obtained by taking a point
 in $\G$ and evolving from the Cauchy surface $E(\Sigma)$ to get a
 solution of $(g^{ab}\nabla_{a}\nabla_{b}-m^{2})\psi =0$. That is,
the specification of a point in $\G$ is appropriate initial data for
determining
 a solution to the equation of motion. The inverse map,
$I_{E}^{-1}:V \to \G$,
 takes a point $\psi \in V$ and finds the Cauchy data induced on
$\Sigma$ by virtue of the embedding $E$: $\varphi=E^{*}\psi$ and
$\pi=E^{*}(\sqrt{h}\pounds_{n}\psi)$, where $\pounds_{n}$ is the
Lie derivative along the normal to the Cauchy surface $E(\Sigma)$
 and $h$ is the determinant of the induced metric on
$E(\Sigma)$.

 It is worth pointing out that $\G$ and $V$ are equipped
  with a (natural) symplectic structure $\w_{\G}$ and $\w_{V}$,
respectively, that provides the space of classical observables,
which are (alternatively) represented by smooth real valued
functions on $\G$ or $V$, with an algebraic structure via the
Poisson bracket. On the space of solutions $V$, the symplectic
structure is
\begin{equation}
\label{ss-v}
\w_{V}(\psi_{1}, \psi_{2})= \int_{E(\Sigma)}\sqrt{h}(\psi_{2}
\pounds_{n}\psi_{1}-\psi_{1}\pounds_{n}\psi_{2})
\end{equation}
while on the space of Cauchy data, $\G$, is given by
\begin{equation}
\label{ss-gamma}
\w_{\G}\bigl((\varphi_{1},\pi_{1}),(\varphi_{2},\pi_{2}) \bigr)=
\int_{\Sigma}(\varphi_{2}\pi_{1}-\varphi_{1}\pi_{2})
\end{equation}
From Eqs(\ref{ss-v})-(\ref{ss-gamma}) and the specification of the
isomorphism $I_{E}$, it is obvious that $\Omega_{\Gamma}=I^{*}_{E}
\Omega_{V}$; i.e., $I_{E}$ is a symplectic map.

Now, let $E_{I}(\Sigma)$ and $E_{F}(\Sigma)$ be any given initial and
final Cauchy surfaces, represented by embeddings $E_{I}$ and $E_{F}$.
{\it{Time evolution}} from $E_{I}(\Sigma)$ to $E_{F}(\Sigma)$ can be
viewed as a bijection $t_{(E_{I},E_{F})}:\G \to \G$ on the space of
Cauchy data \cite{torrvar2}:  $t_{(E_{I},E_{F})}:=
I_{E_{F}}^{-1}\circ I_{E_{I}}$.
 Thus, the recipe is: (a) take initial data on $E_{I}(\Sigma)$, (b)
 evolve it to a solution of the Klein-Gordon equation, and (c) find
 the corresponding pair induced on $E_{F}(\Sigma)$  by this solution.
 Notice that this map also defines {\it{time evolution}} on the space
 of solutions, $V$, through the natural induced bijection
 $T_{(E_{I},E_{F})}:=I_{E_{I}}\circ t_{(E_{I},E_{F})}
 \circ I_{E_{I}}^{-1}$. The three steps of the recipe are now: (a)
take a solution to the field equation, (b) find the data induced on
$E_{F}(\Sigma)$, and (c) take the data as initial data on
$E_{I}(\Sigma)$ and find the resulting solution.  It is
straightforward to see,
 from the embedding independence of (\ref{ss-v}) and from
 (\ref{ss-gamma}), that each transformation is a symplectic
 isomorphism. i.e., $t^{*}_{(E_{I},E_{F})}\w_{\G}=\w_{\G}$
 and $T^{*}_{(E_{I},E_{F})}\w_{V}=\w_{V}$.

For our particular case, we shall construct dynamical evolution between
any two flat Cauchy surfaces $E_{I}(T^{2}):=(T_{I},x^{i})$ and
$E_{F}(T^{2}):=(T_{F},x^{i})$, where $x^{i}=(\theta , \sigma) \in
(0, 2 \pi)$ are coordinates on $T^{2}$ and $T$ is the smooth ``time
coordinate'' on $(M\simeq T^{2}\times {\bf{R}},^{{\mbox{\tiny{(f)}}}}
g_{ab})$ such that each surface of constant $T$ is a
 Cauchy surface.
Let us denote by  $\tilde{\psi}$ the resulting solution from the
 action of $T_{(E_{I},E_{F})}$ on $\psi$. Following the
 prescription, we first have to find the induced data on $E_{F}(T^{2})$:

In general $\varphi_{F}=E^{*}_{F}\psi$ and
$\pi_{F}=E^{*}_{F}(\sqrt{h_{F}}\pounds_{n_{F}} \psi)$,
since in our case $E_{F}(T^{2})=(T_{F},x^{i})$ and $\psi$ depend on the
coordinates $T$ and $\theta$ only, we thus have that
$\varphi_{F}=\psi(\theta,T_{F})$ and $\pi_{F}=[
T\partial_{T}\psi(\theta,T)]\vert_{T=T_{F}}$. Thus, from the
explicit form (\ref{kg-sol}) for solutions of the Klein-Gordon equation,
we have that
\be
\label{varphi-on-E(F)}
\varphi_{F}=\Im (A_{0})+\Re (A_{0})\ln T_{F}+ \sum_{m \neq 0}
\Re[B_{m}H_{0}^{(1)}(\vert m \vert T_{F})]
\ee
\be
\label{pi-on-E(F)}
\pi_{F}= \Re (A_{0})-T_{F}\sum_{m \neq 0}\vert m \vert
\Re[B_{m}H_{1}^{(1)}(\vert m \vert T_{F})]
\ee
where $B_m= A_{m}e^{im \theta}$.

The next step in the prescription is to take the pair
$(\varphi_{F},\pi_{F})$
as initial data on $E_{I}(T^{2})$ and find the resulting solution
$\tilde{\psi}$. That is,  we have to solve for
$\{\tilde{A}_{k},\overline{\tilde{A}}_{k}\}_{k\in {\bf{Z}}}$
the following system
\begin{eqnarray}
\label{sys1}
\psi(\theta,T_{F})=\tilde{\psi}(\theta,T_{I}) \\
\label{sys2}
[T \partial_{T}\psi(\theta,T)]\vert_{T=T_{F}}=
[T\partial_{T}\tilde{\psi}(\theta,T)]\vert_{T=T_{I}}
\end{eqnarray}
where $\tilde{\psi}(\theta,T_{I})=E^{*}_{I}\tilde{\psi}$ and
 $[T\partial_{T}\tilde{\psi}(\theta,T)]\vert_{T=T_{I}}=
E^{*}_{I}(\sqrt{h_{I}}\pounds_{n_{I}} \tilde{\psi})$ are explicitly given by
\be
\label{varphi-on-E(I)}
\tilde{\psi}(\theta,T_{I}) = \Im (\tilde{A}_{0})+\Re
(\tilde{A}_{0})\ln T_{I}+ \sum_{m \neq 0} \Re [\tilde{B}_{m}
H_{0}^{(1)}(\vert m \vert T_{I})]
\ee
\be
\label{pi-on-E(I)}
[T\partial_{T}\tilde{\psi}(\theta,T)]\vert_{T=T_{I}} =
\Re (\tilde{A}_{0})-T_{I}\sum_{m \neq 0}\vert m \vert \Re
[\tilde{B}_{m} H_{1}^{(1)}(\vert m \vert T_{I})]
\ee
with $\tilde{B}_m= \tilde{A}_{m}e^{im \theta}$.

Using the orthogonality property $\oint e^{i(n-m)\theta}=2\pi
 \delta_{n,m}$, the well-known relation
$H_{0}^{(1)}(x)\overline{H_{1}^{(1)}(x)}-H_{1}^{(1)}(x)
\overline{H_{0}^{(1)}(x)}={{4i}\over{\pi x}}$ (where $x>0$)
and the explicit expression for the fields, given by equations
(\ref{varphi-on-E(F)}), (\ref{pi-on-E(F)}), (\ref{varphi-on-E(I)})
 and (\ref{pi-on-E(I)}), it is not
difficult to see that the system (\ref{sys1})-(\ref{sys2}) is
solved by
\be
\Re (\tilde{A}_{0})=\Re (A_{0})
\ee
\be
\Im (\tilde{A}_{0})=\Im(A_{0})+\Re (A_{0}) \ln (T_{F}/T_{I})
\ee
for $k=0$, and
\be
\tilde{A}_{k}={{i \pi}\over{4}}[F(y_{k},x_{k})-
\overline{F(x_{k},y_{k})}]A_{k}+{{i \pi}\over{4}}
[ G(y_{k},x_{k})-G(x_{k},y_{k})]\overline{A_{-k}}
\ee
for all $k \neq 0$, where $x_{k}:=\vert k\vert T_{I}$, $y_{k}:=
\vert k \vert T_{F}$, $F(r,s):=rH_{1}^{(1)}(r)\overline{H_{0}^{(1)}(s)}$
and $G(r,s)=r \overline{H_{1}^{(1)}(r)}\overline{H_{0}^{(1)}(s)}$.

Therefore, the symplectic transformation $T_{(E_{I},E_{F})}$ defines,
and is defined by, a transformation of $\overline{A_{m}}$:
\begin{equation}
\label{coeff-cc-trans}
\overline{\tilde{A}_{k}}=\sum_{l \in {\bf{Z}}} \chi _{kl}A_{l}+
\xi_{kl}\overline{A_{l}}
\end{equation}
where
\be
\label{chis}
\chi _{k0}=-{{i}\over{2}}\ln ( T_{F}/T_{I}) \delta_{k,0} \:, \:\:
\chi _{kl}={{i \pi}\over{4}}[ \overline{G(x_{l},y_{l})}-
\overline{G(y_{l},x_{l})} ]\delta_{l,-k}\: (\forall l \neq 0)
\ee
\be
\xi _{k0}=[1-{{i}\over{2}}\ln ( T_{F}/T_{I})] \delta_{k,0}\: , \:\:
\xi _{kl}={{i \pi}\over{4}}[ F(x_{l},y_{l})- \overline{F(y_{l},x_{l})}]
\delta_{l,k}\: (\forall l \neq 0)
\ee

Obviously $\overline{\tilde{A}_{k}}=\overline{A_{k}}$ for all
$k \in {\bf{Z}}$ when $T_{I}=T_{F}$ (i.e., when $T_{(E_{I},E_{F})}$
is the identity map).

\subsection{Quantum Implementability}
\label{sec:3.b}

The question we want to address in this part is whether or not
classical evolution on the fictitious background is implementable
at the quantum level. A particularly convenient approach to this
issue is given by the algebraic approach of QFT, since the notion
of implementability of symplectic transformations on a Hilbert
space formulation is defined in a natural way \cite{torrvar2}. The
main idea in the algebraic approach is to formulate the quantum
theory in such a way that the observables become the relevant
objects and the quantum states are ``secondary", they are taken to
``act" on operators to produce numbers. The basic ingredients of
this formulation are two, namely : (1) a $C^{*}$-algebra
${\cal{A}}$ of observables, and (2) states $\omega:{\cal{A}}\to
{\bf{C}}$, which are positive
 linear functionals ($\omega(A^{*}A)\geq 0$ $\forall A\in
{\cal{A}}$) such that $\omega({\bf{1}})=1$. The value of the
 state $\omega$ acting on the observable $A$ can be
interpreted as the expectation
value of the operator $A$ on the state $\omega$, i.e.,
$\la A\ra=\omega(A)$.

For free (linear) fields it is possible to construct the Weyl algebra of
quantum abstract operators from the elementary classical observables
(equipped with an algebraic structure given by the Poisson bracket).
The elements of this $C^{*}$-algebra are taken to be the
fundamental observables for the quantum theory, thus the (natural)
algebra ${\cal{A}}$ for free fields is the Weyl algebra. Let $(Y,\w_{Y})$
be a symplectic vector space, each generator $W(y)$ of the Weyl algebra
is the ``exponentiated'' version of the linear observable
$\w_{Y}(y,\,\cdot \,)$. These generators satisfy the Weyl
relations{\footnote{The CCR
that correspond to operators $\hat{\w}_{Y}(y,\,\cdot \,)$ get now
 replaced by the quantum Weyl relations.}}:
\be
W(y)^{*}=W(-y)\:, \:\: W(y_{1})W(y_{2})=
e^{-{{i}\over{2}}\w_{Y}(y_{1},y_{2})}W(y_{1}+y_{2})
\ee

Given a symplectic transformation $f$ on $Y$, there is an
associated $*$-automorphism of ${\cal{A}}$, $\alpha_{f}:{\cal{A}}\to
{\cal{A}}$, defined by $\alpha_{f} \cdot W(y):=W(f[y])$. In particular,
the symplectic transformation $T_{(E_{I},E_{F})}$ representing time
evolution from $E_{I}=(T_{I},x^{i})$ to $E_{F}=(T_{F},x^{i})$
defines the $*$-automorphism
$\alpha_{(E_{I},E_{F})}$. Thus, from the algebraic point of
view, if we assign the state $\omega$
to the initial time as represented by the embedding $E_{I}$, the
expectation value of the observable $W\in {\cal{A}}$ on $E_{I}$ is
given by
\be
\la W\ra_{E_{I}}=\omega(W)
\ee
Let us consider the Weyl
generator $W(\psi)$ labeled by $\psi$. Under the symplectic
transformation $T_{(E_{I},E_{F})}$, the label goes to
$\tilde{\psi}$ and the relation between $W(\psi)$ and
$W(\tilde{\psi})$ is given by
$\alpha_{(E_{I},E_{F})}W(\psi)=W(\tilde{\psi})$. Since
$T_{(E_{I},E_{F})}$ dictates time evolution at classical level,
one can interpret the change $W(\psi)\to
\alpha_{(E_{I},E_{F})}W(\psi)$ as a counterpart in the
observables. That is, $W(\psi)\to \alpha_{(E_{I},E_{F})}W(\psi)$
is the mathematical representation of time evolution of
observables in the Heisenberg picture. Thus, while in the
Heisenberg picture the expectation value of the observable
$W(\psi)$ at final time is given by $\la W(\psi)\ra_{E_{F}}=
\omega(\alpha_{(E_{I},E_{F})} \cdot W(\psi))$, in the
Schr\"{o}dinger picture is $\la W(\psi)\ra_{E_{F}}=\omega_{E_{F}}(W(\psi))$.
Therefore, the final state $\omega_{E_{F}}$, obtained by evolving
the initial state $\omega$, is given by
$\omega \circ\alpha_{(E_{I},E_{F})}$.

Now, in order to know if time evolution between any two flat
Cauchy surfaces is well defined in the framework of the Hilbert
space formulation, we have to introduce the GNS construction that
tells us how the quantization in the old sense (that is, a
representation of the Weyl relations on a Hilbert space) and the
algebraic approach are related:

{\it{Let ${\cal A}$ be a $C^*$-algebra with unit and let $\omega:{\cal
A}\rightarrow \C$  be a state. Then there exist a Hilbert space
${\cal H}$, a representation $\pi:{\cal A}\rightarrow L({\cal H})$
and a vector $|\Psi_0\ra\in {\cal H}$ such that, $ \omega(A)=\la
\Psi_0,\pi(A)\Psi_0\ra_{\cal H} \label{teo-expect}$. Furthermore, the vector
$|\Psi_0\ra$ is cyclic. The triplet $({\cal H},\pi,|\Psi_0\ra)$
with these properties is unique (up to unitary equivalence)}}.

With this in hand, we have a precise way to `go down' transformations
on the $C^*$-algebra to a given Hilbert space representation. Thus, a
symplectic transformation $f:Y\to Y$, with corresponding algebra
automorphism $\alpha_{f}:{\cal {A}}\to {\cal {A}}$, is unitarily
implementable \cite{torrvar2} if there is a unitary transformation
$U:{\cal H} \to {\cal H}$ on the Hilbert space ${\cal H}$ such that,
for any $W\in {\cal A}$, $U^{-1}\pi(W)U=\pi(\alpha_{f}\cdot W)$.

Because $\omega$ and its transform $\omega \circ \alpha_{f}$ will not always
define unitarily equivalent Hilbert space representations, thus not all
symplectic transformations $f$ will be implementable in field theory.
In our case, we are interested on the implementability of
$f=T_{(E_{I},E_{F})}$ on the symmetric Fock space $\bar{\cal{F}}$,
constructed from the so-called ``one-particle'' Hilbert space,
${\cal {H}}_{0}$, which elements are the complex functions
\be
\label{1phs}
\Psi= \sum_{m \in {\bf{Z}}} \overline{f_{m}(\theta,T)}\: \overline{A_{m}}
\ee
determined by the natural splitting of $V_{{\bf{C}}}$, the
complexification of $V$, on negative and {\it{positive}} parts
through the complex structure $J_{V}$.

The continuous{\footnote{Actually, it can be shown that there is a constant
$b$ such that, for all $\psi \in V$, $\vert \vert T_{(E_{I},E_{F})}
\psi \vert \vert  \leq b \vert \vert \psi \vert \vert$ in the norm
$\vert \vert \psi \vert \vert ^{2}= \w_{V}(J_{V}\psi,\psi)$. }}
transformation (\ref{coeff-cc-trans}) defines a pair of bounded linear
maps $\xi:{\cal{H}}_{0} \to {\cal{H}}_{0}$ and $\chi:
{\cal{H}}_{0} \to \overline{{\cal{H}}_{0}}$, where
$\overline{{\cal{H}}_{0}}$ is the complex conjugate
space to ${\cal{H}}_{0}$. With $\Psi$ given by (\ref{1phs}), we have
\be
\xi \cdot \Psi =\sum _{m,l \in {\bf{Z}}}
\overline{f_{m}(\theta,T)} \xi_{ml}\:\overline{A_{l}}
\ee
and
\be
\chi \cdot \Psi = \sum _{m,l \in {\bf{Z}}} f_{m}(\theta,T)
\overline{\chi_{ml}}\: \overline{A_{l}}
\ee
The automorphism $\alpha_{(E_{I},E_{F})}$ associated with
$T_{(E_{I},E_{F})}$ is unitarily implementable with respect
to the Fock representation $(\bar{\cal{F}}={\cal{F}}_{s}
({\cal{H}}_{0}),\pi)$
if and only if the operator $\chi$ is Hilbert-Schmidt \cite{honegger}.
i.e., iff
\be
\label{hs-condition}
\sum _{m,l \in {\bf{Z}}}  \vert \chi _{lm} \vert ^{2} < \infty
\ee

Since $\sum _{m,l \in {\bf{Z}}}  \vert \chi _{lm} \vert ^{2}=
\vert {{1}\over{2}} \ln (T_{F}/T_{I})\vert ^{2}+\sum_{l,m \neq 0}
\vert \chi _{lm} \vert ^{2}$, then according
to Eq.(\ref{chis})
condition (\ref{hs-condition}) is equivalent to
\be
\label{h-s-condition}
\sum _{m \neq 0}\bigl(\Re [g_{m}(x_{m},y_{m})]\bigr)^{2}
< \infty \:\:\:\: \mbox{and }\:\:\:\: \sum _{m \neq 0}
\bigl(\Im [g_{m}(x_{m},y_{m})]\bigr)^{2}< \infty
\ee
where $g_{m}(r,s):=\overline{G(r,s)}-\overline{G(s,r)}$.
Using the definition of Hankel function in terms of Bessel
and Neumann functions (for $n=0$ or $1$), the first
 condition in (\ref{h-s-condition}) can be written as follows
\begin{equation}
\label{lambda-cond}
\sum_{m=1}^{\infty}\bigl( \Lambda_{m}[a,y_{m}]\bigr)^{2}< \infty
\end{equation}
where $\Lambda_{m}[a,y_{m}]:=m[a(J_{0}(y_{m})J_{1}(ay_{m})-
N_{0}(y_{m})N_{1}(ay_{m}))+N_{1}(y_{m})N_{0}(ay_{m})-J_{1}(y_{m})
J_{0}(ay_{m})]$ and $a:=T_{I}/T_{F}$. In the asymptotic region
$x\gg 1$ (for $n=0$ or $1$) the behavior of Bessel and Neumann
functions is given by $J_{n}(x)\approx \sqrt{{2}\over{\pi x}}\cos
\bigl(x-\bigl( n+{{1}\over{2}}\bigr){{\pi}\over{2}} \bigr)$ and
$N_{n}(x)\approx \sqrt{{2}\over{\pi x}}\sin \bigl(x-\bigl(
n+{{1}\over{2}}\bigr){{\pi}\over{2}} \bigr)$ respectively, then
$\Lambda_{m}[a,y_{m}]\approx {{2m}\over{\sqrt{a}
\pi y_{m}}}(a-1)\cos(y_{m}(1+a)-\pi)$ for $m\gg 1$
is the asymptotic behavior of $\Lambda_{m}[a,y_{m}]$
 and (\ref{lambda-cond}) becomes
\begin{equation}
\label{cos-cond}
{{4(1-a)^{2}}\over{a\pi^{2}T_{F}^{2}}}
\sum_{m=N}^{\infty}\cos^{2}(T_{F}(1+a)m) < \infty
\end{equation}
where $N\gg 1$. Notice that for the case $a=1$ this
condition is trivially satisfied, as expected.

Thus, unitary implementability implies that (\ref{cos-cond})
is satisfied
for all $a \in (0,1)$ and hence, if there are particular values
of $T_{F}$ ($r_{0}>0$) and $a$
 ($a_{0}\in (0,1)$) such that $\sum_{m=N}^{\infty}\cos^{2}(r_{0}(1+a_{0})m)$
 diverges, then $\alpha_{(E_{I},E_{F})}$ will not be unitarily implemented.
 In particular, by choosing $T_{F}={{\pi}\over{1+a}}$ every integer $m\geq N$
  corresponds to a maximum of $\cos^{2}(T_{F}(1+a)x)$ and therefore the sum
  in (\ref{cos-cond}) diverges. Thus, the transformation associated with
  $T_{(E_{I},E_{F})}$ is {\it{not}} unitarily implementable with respect
  to the Fock representation $({\cal{F}}_{s}({\cal{H}}_{0}),\pi)$ and hence
  classical time evolution, dictated by $T_{(E_{I},E_{F})}$, does not have a
  quantum analog in the Hilbert space formulation via a unitary
operator. In this case, the Schr\"{o}dinger picture is not
available to describe functional evolution using the Fock space
representation of the quantum theory, consequently the
``Schr\"{o}dinger equation" associated with the Hamilton
operator $\hat{H}(T)$ can not be interpreted as an evolution
equation (on the fictitious background) for quantum states.

\section{Discussion and Conclusions}
\label{sec:5}

In this work we have analyzed the quantization of polarized Gowdy
$T^3$ cosmological models as carried out by Pierri.  We have found
explicitly the symplectic transformation that determines the
classical dynamical evolution $T_{(E_I, E_F)}$ given by the phase
space function $T$. We have shown that this symplectic
transformation does not have a quantum analog in the Hilbert space
formulation via a unitary operator. This means that the classical
dynamics of Gowdy $T^3$ cosmological models cannot be implemented
in the context of Pierri's quantization procedure.
Let us now discuss the implications of this negative result for
two related areas, namely for canonical quantum gravity and for
quantization of fields on curved manifolds.

\vskip 0.2cm
\noindent
 {\it Canonical Quantum Gravity.}

First of all let us recall that
in canonical quantum gravity, the theory is defined over an
``abstract'' manifold $\Sigma$ which in our case is given by
$\Sigma=T^2$. There is no spacetime and therefore no notion of
an embedding of $\Sigma$ into this spacetime. What we have is, in
the gauge fixed scenario, a reduced phase space representing the
true degrees of freedom, and, in the quantum theory, a Hilbert space
of physical states and physical observables defined on it. This is
the frozen formalism description. When the deparametrization procedure
was introduced classically, an artificial notion of time evolution was
created that allows to ``evolve'' any set of (physical) initial
conditions on $\Sigma$ into a one parameter family of initial conditions
with a precise spacetime interpretation. This one parameter family
is generated via a canonical transformation generated by the reduced
Hamiltonian. In quantum gravity, however, the parameter $T$ can not be
thought, a priory, as a a time function in a spacetime for the only
reason that a spacetime notion is absent. In which sense is then
useful the parameter $T$?   Recall from the discussion in Sec.~\ref{sec:3}
that the notion of time evolution in the algebraic formulation is well
defined, giving rise to the Heisenberg picture:  We have a unique state
$|\Psi\rangle_{T_0}$, defined on a preferred and fixed $\Sigma_0$, and
operators acting on it which could be ``time dependent''. This is the
place where the parameter $T$ plays a central role. We can have,
for instance, a one parameter family of observables, say $\hat{V}_T$,
corresponding to ``the volume of the Universe at time $T$''
\cite{pierri2}. In the
standard formulation of quantum theory, where a unitary evolution
operator $\hat{U}(T,T_0)$ exists, we can relate the operators belonging
to the family via unitary transformations. One can also construct the
Schr\"odinger picture and have a one-parameter family of states
that ``evolve'' in time $T$, using the standard construction. Canonical
quantum gravity is most naturally constructed in the Heisenberg picture,
where the physical operators correspond to the so called ``evolving constants
of the motion''. Therefore, any operator ${\cal O}_T$  labeled by the
time $T$, if it is well defined (i.e. if it leaves the Hilbert space, or
a dense subset of it, invariant), can be regarded as a physically
meaningful object representing the classical observable at ``time $T$''.
What is lost  in the absence of the operator
 $\hat{U}(T,T_0)$, as is our case, is the ``unitary equivalence'' of the
operators  ${\cal O}_T$ for all values of $T$. In this sense, unitary
time evolution and the Schr\"odinger picture are lost. Unitary evolution
is one of the pillars of present quantum theory, and theories
that do not satisfy this property suffer from the rejection of the
community, since the theory becomes unable to make predictions due to
the lack of conservation of probability.
This would be the case, for instance, in the event of the evaporation
of a black hole via Hawking radiation. Physicists have always tried
to avoid such descriptions and look for explanations that are ``unitary''.
However, as we would like to argue, canonical quantum gravity is conceptually
very different from the standard description of quantum theory with a
preferred and external Newtonian time, so one should look for
more involved arguments before dismissing a particular theory.
In the Hamiltonian description,
time evolution is pure gauge, so strictly speaking one should only
meaningfully discuss physical observables on the reduced phase space.
There is no time evolution and no dynamics. Any deparametrization is
classically equivalent, giving rise to fictitious dynamics via canonical
transformations. There is no compelling reason to expect that there
is a preferred deparametrization that will be meaningful
quantum-mechanically. Thus, there is no logical contradiction to the
result that a particular choice can not be unitarily implemented.

Within this perspective, the
parameter $T$ that was introduced artificially at the classical level
bears  no fundamental physical significance. The fact that the quantum
theory does not endorse this choice should not be enough reason to
dismiss it.  However, it should be clear that
 the absence of unitary transformation reduces
significantly the importance of this quantization, since the
Heisenberg operators
${\cal O}_T$ are not well defined. That is, their spectra, expectation values,
etc., depend on the choice of the value of $T_0$. Had we chosen a different
value  $T=T^\prime_0$ and therefore different Heisenberg
state $|\Psi\ra_{T^\prime_0}$,
we would get different operators ${\cal O}_T^\prime\neq
{\cal O}_T$ (for the same value of $T$). 
A minimum requirement for the consistency of the
quantization is that the operators ${\cal O}_T^\prime$ and ${\cal O}_T$
be (unitary) equivalent. Thus, the quantization is physically unacceptable.

However, it
is not completely clear whether this negative result holds for any
choice of a set of creation and annihilation operators (or,
equivalently, choice of complex structure $J$). The original
choice in Ref.~\cite{pierri1} seems natural from the viewpoint
of the explicit form of the
solutions of the Klein-Gordon equation, and the fact that is time
independent and therefore there is no ``particle creation''.
However, further work
is needed in order to understand whether there exist different
choices of $J$ and therefore of representations of the CCR for which
``time evolution'' is a well defined concept.
Unitary implementability
might even be a criteria leading to a physically relevant
quantization.

\vskip 0.2cm
\noindent
{\it Quantum Fields on Curved Surfaces.}

The issue of formulating time evolution between arbitrary
Cauchy surfaces in the quantum
theory of fields goes back to the work of Dirac \cite{dirac}.
However, it is only recently that unitary implementability of
arbitrary time evolution
has been considered. Somewhat surprisingly,
it has been recognized that even
for free fields on Minkowski spacetime, time evolution between
arbitrary Cauchy surfaces is not unitarily implementable in three
and higher spacetime dimensions \cite{torrvar2}. The failure is in
general attributed to the fact that time evolution between
arbitrary surfaces is not generated by an isometry of the
background metric \cite{helfer,torrvar2}. It is also known that in
two dimensions, for the standard quantization coming from the
symmetries of the system, time evolution is well defined for
arbitrary Cauchy surfaces with topology of a circle. However, for
our model, even when it is a truly
 two dimensional model ($\psi$ depends only on
$\theta$ and $T$), it does not satisfy a free scalar equation (it
is instead related to a Liouville model). Therefore, there is no
contradiction with the fact that time evolution is not unitary.
From the three dimensional perspective, the theory is given by a
free scalar field on a flat background, but in which the vector
field   $\partial/\partial T$ that generates the natural time
evolution is not an isometry of the background spacetime. Thus, it
is interesting to see that in this case, even for the simplest
Cauchy surfaces (flat and parallel in the given chart), time
evolution is not implementable. It is also interesting to note
that particle creation and non-unitary time evolution do not imply
each other, as noted in \cite{torrvar2}. As previously mentioned,
it is not clear whether different representations of the CCR would
yield unitary quantum theories. Namely, is there a choice of $J$
that will render the theory unitary? Would it be unique? We shall
leave these questions for future investigations.

\vskip 0.2cm
\noindent
Note added: After submitting this paper, we learned that similar
results were independently found by Torre \cite{torr3}.

\section*{Acknowledgments}
We would like to thank the referee for helpful comments and C. Torre
for correspondence.
This work has been supported by DGAPA--UNAM grant No. IN-112401  and
CONACYT grants 36581-E and J32754-E. This work was also partially
supported by NSF grant  No. PHY-0010061. J.C. was supported by a
CONACYT-UNAM(DGEP) Graduate Fellowship.


\begin{thebibliography}{99}

\bibitem{misner1} C.~W.~Misner, In {\em Magic without Magic: John
Archibald Wheeler}, J. Klauder (Ed.) (Freeman, San Francisco,
1972).



\bibitem{torre} C.~G.~Torre,
``Midisuperspace models of canonical quantum gravity,''
Int.\ J.\ Theor.\ Phys.\  {\bf 38}, 1081 (1999).

\bibitem{ashpie} A. Ashtekar and M. Pierri,
``Probing quantum gravity through exactly soluble midi-superspaces. I,''
J.\ Math.\ Phys.\  {\bf 37}, 6250 (1996).


\bibitem{pierri1}
M. Pierri,
``Probing quantum general relativity through exactly soluble
midi-superspaces. II: Polarized Gowdy models,''
Int.\ J.\ Mod.\ Phys.\ D {\bf 11}, 135 (2002).


\bibitem{pierri2} M. Pierri,
``Hamiltonian and volume operators,''
arXiv:gr-qc/0201013.

\bibitem{gowdy} R.~H. Gowdy,
``Gravitational Waves In Closed Universes,''
Phys.\ Rev.\ Lett.\  {\bf 27}, 826 (1971);
Ann. Phys. {\bf 83}, 203 (1974).

\bibitem{berger} B.~K.~Berger, Ann. Phys. {\bf 83}, 458 (1974);
``Quantum Cosmology: Exact Solution For The Gowdy $T^3$ Model,''
Phys.\ Rev.\ D {\bf 11}, 2770 (1975).


\bibitem{misner2} C.~W.~Misner, ``A minisuperspace example:
The Gowdy $T^3$ cosmology'', Phys. Rev. {\bf D8}, 3271 (1973).

\bibitem{husain} V. Husain, ``Quantum Effects on the
syngularity of the Gowdy Cosmology", Class. Quantum Grav. {\bf 5},
1587 (1987).

\bibitem{mena} G.~A.~Mena Marugan,
``Canonical quantization of the Gowdy model,'' Phys.\ Rev.\ D {\bf
56}, 908 (1997).


\bibitem{kuchar}
K.~Kucha\u{r},
``Dirac Constraint Quantization of a Parametrized Field Theory by
Anomaly - Free Operator Representations of Space-Time Diffeomorphisms,''
Phys.\ Rev.\ D {\bf 39}, 2263 (1989);
K.~Kuchar,
``Parametrized Scalar Field on $R \times S(1)$: Dynamical Pictures,
Space-Time Diffeomorphisms, and Conformal Isometries,''
Phys.\ Rev.\ D {\bf 39}, 1579 (1989).


\bibitem{helfer} A.~D.~Helfer,
``The Stress-Energy Operator,''
Class.\ Quant.\ Grav.\  {\bf 13}, L129 (1996)
[arXiv:gr-qc/9602060].


\bibitem{torrvar1}
C. G. Torre and M. Varadarajan, ``Quantum fields at any time,''
Phys.\ Rev.\ D {\bf 58}, 064007 (1998).

\bibitem{torrvar2} C.~G.~Torre and M.~Varadarajan,
``Functional evolution of free quantum fields,''
Class.\ Quant.\ Grav.\  {\bf 16}, 2651 (1999).


\bibitem{birdav} N. D. Birrel and  P. C. W. Davies, {\it Quantum fields
in curved space} (Cambridge University Press, Cambridge, 1982).

\bibitem{Ash-Bomb-Reul-wald-books}
A.Ashtekar, L.Bombelli and O.Reula, in {\it{Mechanics, Analysis and
Geometry: 200 Years After Lagrange}} (North-Holland, New York 1991);
R.M.Wald, {\it{Quantum Field Theory in Curved Spacetime and Black Hole
Thermodynamics}} (University of Chicago Press, Chicago, 1994).

\bibitem{honegger}
R.Honegger and A.Rieckers, ``Squeezing Bogoliubov transformations
on the infinite mode CCR-algebra'',
J.\ Math.\ Phys. {\bf 37}, 4291 (1996).

\bibitem{dirac} P.A.M. Dirac, {\it Lectures on Quantum Mechanics},
(New York, Yeshiva University, 1964).

\bibitem{arfken} G.B. Arfken and H.J. Weber, {\it Mathematical Methods 
for Physicists}, (Academic Press, 1995).

\bibitem{torr3} C.~G.~Torre, private communication.

\end{thebibliography}
\end{document}